\documentclass[fleqn,10pt]{wlscirep}
\usepackage[utf8]{inputenc}
\usepackage[T1]{fontenc}

\title{Rigorous analysis of the topologically protected edge states in the 
quantum spin Hall phase of the armchair ribbon geometry}

\author[1]{Mozhgan Sadeghizadeh}
\author[1,+]{Morteza Soltani}
\author[1,*]{Mohsen Amini}
\affil[1]{Department of Physics, Faculty of Physics, University of Isfahan, Isfahan 81746-73441, Iran}

\affil[+]{mo.soltani@sci.ui.ac.ir}
\affil[*]{msn.amini@sci.ui.ac.ir}

\begin{abstract}
Studying the edge states of a topological system and extracting their topological properties is of great importance in understanding and characterizing these systems. 
In this paper, we present a novel analytical approach for obtaining explicit expressions for the edge states in the Kane-Mele model within a ribbon geometry featuring armchair boundaries. Our approach involves a mapping procedure that transforms the system into an extended Su-Schrieffer-Heeger model, specifically a two-leg ladder, in momentum space. Through rigorous derivation, we determine various analytical properties of the edge states, including their wave functions and energy dispersion. Additionally, we investigate the condition for topological transition by solely analyzing the edge states, and we elucidate the underlying reasons for the violation of the bulk-edge correspondence in relatively narrow ribbons. Our findings shed light on the unique characteristics of the edge states in the quantum spin Hall phase of the Kane-Mele model and provide valuable insights into the topological properties of such systems.

\end{abstract}
\begin{document}

\flushbottom
\maketitle

\thispagestyle{empty}

\section*{Introduction}

The quantum spin Hall (QSH) effect~\cite{QSH1, QSH2} is a fascinating phenomenon in condensed matter physics, which manifests as a gapless edge state in a two-dimensional topological insulator that is protected by time-reversal symmetry. 
The Kane-Mele model~\cite{Kane1,Kane2} is one of the simplest model Hamiltonians that exhibits the QSH effect, and has attracted a lot of attention in recent years due to its potential applications in spintronics and quantum computing~\cite{He, Vaezi, Nagaosa, Amini1}.
The emergence of such edge states within the energy band gap in the QSH systems is intimately connected to the existence of nontrivial topological properties in their bulk bands which are referred to as topological invariants.
These topological invariants can be associated with a nonzero topological quantum number 
and due to the bulk-edge correspondence~\cite{Bulk-Edge} one naturally  expects the existence of protected edge states.
These edge states are typically localized at the boundaries or interfaces of the material and are protected from backscattering due to the underlying topological properties.

Typically, the bulk-edge correspondence serves as a valuable tool to comprehend various characteristics of topologically protected edge states by examining the bulk band structure. 
For instance, the existence and stability of edge states for various systems like the odd-parity superconductors, periodically driven quantum systems, non-Hermitian systems, two-dimensional  fermions of high spin, and SU(3) fermions, is investigated by studying  topological properties of the bulk in Refs.~\cite{Sato, Demler, Esaki, Sun, Yau}. 
Nonetheless, it remains intriguing to explore the extent to which information regarding the system's topological properties and transitions can be directly inferred through the study of these edge states. Indeed, understanding the properties of edge states allows researchers to  identify and classify different topological phases and also provides a way to manipulate and control their properties~\cite{ED1, ED2, ED3, Small, ED4, Waka, Rosa, Kimura, Ezawa}. 
In this regard, from theoretical point of view, the availability of explicit expressions allows for a detailed examination of the edge states' characteristics, such as their energy dispersion, spatial distribution, and symmetry properties.

Therefore, it is worth analyzing  the topological phase transition by rigorous derivation of the edge states in a QSH system.
 For this purpose, here, we investigate the topological properties of an armchair honeycomb ribbon by considering the Kane-Mele model~\cite{Kane1,Kane2}. This model is known to exhibit edge states within the energy gap, making it an ideal system for exploring the characteristics and behaviors of these edge states. 
The explicit form of the edge states in the considered model remained elusive until very recently that the successful determination of the edge state wave functions for the zigzag ribbon configuration is reported~\cite{Amini2}. In this study, by focusing on the armchair geometry, we specifically target the unique features associated with the edge states in this type of termination. Through our analysis, we aim to shed light on the fundamental aspects of the armchair edge states and their differences with the zigzag edges states. 
In contrast to the zigzag termination ribbons in which the edge states obtained perturbatively, the edge states of the armchair-terminated ribbons can be obtained exactly.
This breakthrough provides us with a valuable opportunity to delve into the detailed analysis of these edge states and their intriguing properties.
By leveraging this newfound understanding, we aim to identify  the topological transition point  as well as the origin of the bulk-edge correspondence violation for relatively narrow ribbons~\cite{Kondo}.

\section*{Extended SSH model for the Kane-Mele Hamiltonian of an armchair ribbon}\label{SECII}

\begin{figure}
\begin{center}
\includegraphics[width=.9\linewidth]{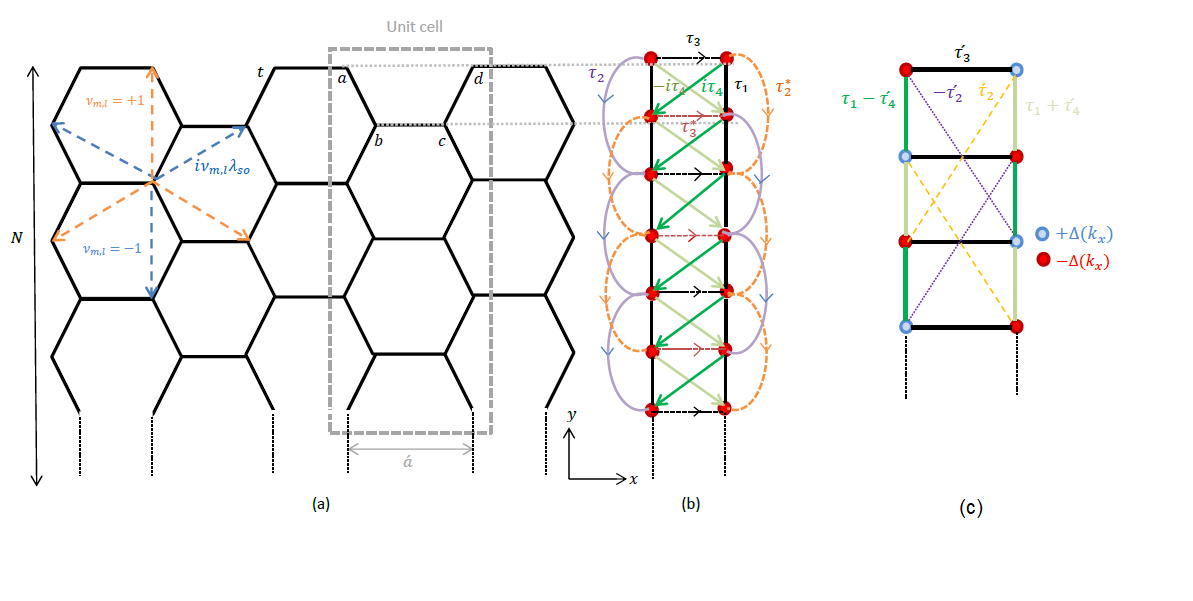} 

\caption{(a) Sketch of the lattice structure of an armchair honeycomb ribbon with width $N$. The sign of the hopping parameter $\nu$ is indicated by dashed lines.
(b) Schematic representation of the corresponding two-leg ladder model obtained after performing the Fourier transformation along the horizontal direction. The new momentum-dependent hopping parameters are represented and defined as $\tau_1=t, \tau_2=i\lambda_{so}, \tau_3=te^{ik_x/2}, \tau_4=2\lambda_{so}\cos(k_x/2)$.
(c) Schematic representation of the transformed two-leg ladder model after applying a proper unitary transformation described in Eq.~\eqref{eq15}. The transformed hopping parameters are denoted with prime notation. The on-site energies $\pm\Delta=\pm t\sin(k_x/2)$ are indicated.
This figure illustrates the step-by-step transformation of the lattice structure and hopping parameters, leading to an equivalent two-leg ladder model in the momentum space. 
}
\label{F1}
\end{center}
\end{figure}

In this section, we delve into the analysis of a Kane–Mele model for the honeycomb lattice with armchair edges (as shown in Fig.~\ref{F1}(a)), specifically focusing on the introduction of a mapping technique that enables us to derive an extended Su-Schrieffer-Heeger (SSH) model for a two-leg ladder structure in the momentum space. This mapping is of significant importance as it provides a framework for investigating the properties of the system, including the emergence of edge states. Prior to delving into the intricacies of the mapping technique, we provide a concise overview of the fundamental aspects of the Kane-Mele model. 
The Hamiltonian that characterizes the dynamics of the Kane-Mele model is expressed as follows~\cite{Kane1,Kane2}:
\begin{eqnarray}\label{eq1}
H=t\sum_{\langle m,l \rangle}{c}_m^\dagger {c}_l+i\lambda_{so}\sum_{\langle\langle m,l \rangle\rangle}\nu_{m,l}{c}_m^\dagger\hat{s}_z {c}_l+i\lambda_{R} \sum_{\langle m,l \rangle}{c}_m^\dagger (s\times d_{m,l})_z{c}_l+M\sum_{m}\epsilon_mc_m^{\dagger}c_m,
\end{eqnarray}
where $ {c}_{m}^{\dagger}=\left( \begin{matrix}{c}_{m\uparrow }^{\dagger}  \\{c}_{m\downarrow }^{\dagger}  \\\end{matrix} \right)$ denotes the creation operator for an electron with spin up/down at the $m$-th site on the honeycomb lattice and $\left\langle m,l \right\rangle $ and $ \left\langle \left\langle m,l \right\rangle  \right\rangle $ stands for summation over the  nearest-neighbor and the next-nearest-neighbor pairs of sites, respectively. 
The first term accounts for the nearest-neighbor hopping with an amplitude of $t$ and the second term captures the spin-orbit couplings with a hopping amplitude denoted as $\lambda_{SO}$ in which $\widehat{{s}_{z}}$ is the $z$-pauli matrix and $ {\nu }_{m,l}$ takes $ \pm 1$ depending on $m$ and $l$.
The third term is the Rashba term with coupling $ {\lambda }_{R}$ that  arises from the interaction between the electron's spin and its momentum in the presence of an electric field gradient. Here, ${d}_{m,l}$ is the unit vector along the nearest neighbor bond pointing from $m$ to $l$. 
The last term represents a staggered on-site potential of strength $M$ in which the parameter  $\epsilon_m=\pm1$ depending on the sub-lattice index $m$.

In the following analysis, we consider a honeycomb lattice with a finite width of $N$ (the number of armchair chains) in the $y$ direction, while maintaining periodic boundary conditions in the $x$ direction, as depicted in Fig.~\ref{F1}. For the sake of simplicity and obtaining analytical results, we omit the third term in this Hamiltonian due to its relatively small influence.
In the absence of the Rashba term and at a finite $M$, the system is in the gapped phase without zero energy states when $\lambda_{SO}=0$.
The introduction of a small $\lambda_{SO}$ initially has no significant impact on the situation. However, the gap undergoes a closure and the system exhibits the QSH phase when the magnitude of the intrinsic spin-orbit coupling parameter  crosses a critical value of $M / 3\sqrt{3}$~\cite{Kane1,Kane2}.
This critical condition, $M < 3\sqrt{3} \lambda_{SO}$, signifies a threshold for the strength of the intrinsic spin-orbit coupling, beyond which the QSH phase emerges.
Hence, we initially set $M=0$; however, we will subsequently examine the case where $M\neq 0$ to investigate the occurrence of a phase transition.

\begin{figure}
\begin{center}
\includegraphics[width=.5\linewidth]{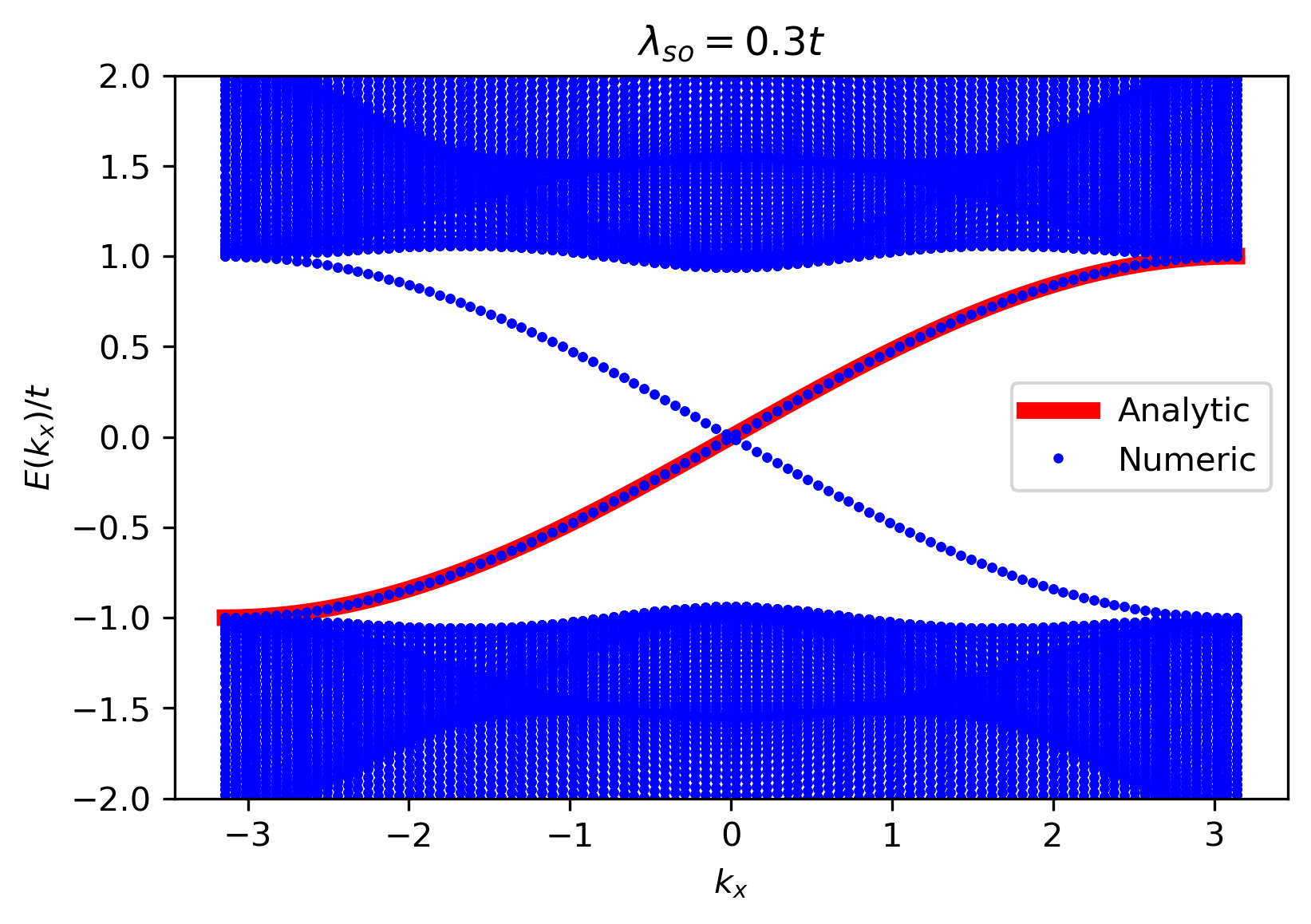} 
\caption{ Numerical energy bands for the Kane-Mele model in a ribbon with armchair boundary plotted as a function of $k_x$, the wave vector parallel to the armchair direction. The spin-orbit coupling strength is set to $\lambda_{SO}=0.3t$ and $M=0$. Additionally, we include the analytical expression obtained in Eq.~\eqref{Spectrum} for the energy spectrum of the edge states in the same system, demonstrating an excellent agreement between the numerical and analytical results. The plot showcases the presence of well-defined energy bands and validates the accuracy of the analytical model in capturing the behavior of the edge states.}
\label{F2}
\end{center}
\end{figure}

To provide a preliminary understanding, it is informative to numerically examine the band structure of ribbons with armchair termination prior to delving into further analysis.
Fig.~\ref{F2} represents the energy spectrum $E(k_x)$ where $k_x$ is the momentum along the ribbon edges for an armchair ribbon with a width of $N=50$ and a spin-orbit coupling strength of $\lambda_{SO}=0.3 t$ in the absence of staggered potential ($M=0$). 
Let us consider the  conventional unit cell as shown in Fig.~\ref{F1} and choose  the unit-cell length to be $a = 1$ to ensure simplicity.
As depicted in Fig.~\ref{F2}, the bulk band-structure displays a distinct gap, while concurrently exhibiting the presence of gapless edge states. This observation aligns precisely with our expectations for the QSH phase.

To initiate our mapping procedure, let us rewrite the Hamiltonian of Eq.~\eqref{eq1} as the following: 
\begin{eqnarray}\label{eq6}
H=\sum_{j}H_{j,j}+H_{j,j+1}+H_{j-1,j},
\end{eqnarray}
where $H_{j,j}$ represents the coupling matrix within the $j$-th unit cell, while $H_{j,j+1}$ refers to the coupling between two adjacent unit cells $j$ and $j+1$ and can be defined as \begin{eqnarray}\label{eq7}
H_{j,j}&=&\sum_{i=1}^{N}t(a_{i,j}^{\dagger}b_{i,j}+b_{i,j}^{\dagger}c_{i,j}+c_{i,j}^{\dagger}d_{i,j}+b_{i,j}^{\dagger}a_{i+1,j}+c_{i,j}^{\dagger}d_{i+1,j})\\ \nonumber
&&+\sum_{i=1}^{N} \lambda_{so}(a_{i,j}^{\dagger}c_{i,j}+a_{i,j}^{\dagger}a_{i+1,j}+b_{i,j}^{\dagger}d_{i,j}+b_{i,j}^{\dagger}b_{i+1,j}+b_{i,j}^{\dagger}d_{i+1,j}+c_{i,j}^{\dagger}a_{i+1,j}+c_{i,j}^{\dagger}c_{i+1,j}+d_{i,j}^{\dagger}d_{i+1,j})+H.C,
\end{eqnarray}
and
\begin{equation}\label{eq8}
H_{j,j+1}=\sum_{i=1}^{N}t(d_{i,j}^{\dagger}a_{i,j+1})+\lambda_{so}(d_{i,j}^{\dagger}b_{i,j+1}+c_{i,j}^{\dagger}a_{i,j+1}+c_{i,j}^{\dagger}a_{i+1,j+1}+d_{i+1,j}^{\dagger}b_{i,j+1}),
\end{equation}
where H.C stands for the hermitian conjugate and it is obvious that 
$H_{j,j-1}=H_{j,j+1}^{\dagger}$.
In this notation, we label the four distinct sites in the unit cell for convenience as $a, b, c,$ and $d$ (as illustrated in Fig.~\ref{F1} (a)) and index $i$ runs over all the armchair chains in the unit cell.
Due to the translational invariant in the $x$-direction,  we can introduce the following Fourier transformation of the fermion annihilation operators of the four basis of the unit cell as the following,
\begin{eqnarray}\label{eq2}
 a(b)_{i,k_x}&=&\sum_{j}e^{i{k_x}j}a(b)_{i,j} , \nonumber\\
 c(d)_{i,k_x}&=&\sum_{j}e^{i{k_x}(j+\frac{1}{2})}c(d)_{i,j},
\end{eqnarray}
in which to get rid of the extra phase factor in the definition of $c(d)$ operators we have used the proper gauge transformation~\cite{Book}. 
Inserting Eq.~\eqref{eq2} into Eqs.~\eqref{eq7} and~\eqref{eq8} one can easily obtain the $k$-space Hamiltonian as
\begin{eqnarray}\label{eq10}
H_{k_x}&=&t \sum_{i=1}^N [a_i^{\dagger}b_i+d_i^{\dagger}c_i+d_i^{\dagger}a_ie^{-ik_x/2}+c_i^{\dagger}b_ie^{ik_x/2}+b_i^{\dagger}a_{i+1}+c_i^{\dagger}d_{i+1}]\\ \nonumber
&&+2i\lambda_{so}cos(k_x/2) \sum_{i=1}^N [d_i^{\dagger}b_i-a_i^{\dagger}c_i-c_i^{\dagger}a_{i+1}+b_i^{\dagger}d_{i+1}]+i\lambda_{so}\sum_{i=1}^N[a_i^{\dagger}a_{i+1}-d_i^{\dagger}d_{i+1}-b_i^{\dagger}b_{i+1}+c_i^{\dagger}c_{i+1}]+H.C.
\end{eqnarray}

It becomes evident that for every value of momentum $k_x$, the Hamiltonian corresponds to an extended SSH model for a two-leg ladder system, as depicted in Fig.~\ref{F1}(b). To enhance convenience, we can now utilize the Pauli matrices $\hat{\sigma}_{\lambda}$ (where $\lambda=x,y,z$) along with the identity operator $\hat{I}_{\sigma}$ to rewrite  the aforementioned Hamiltonian in the following manner:
\begin{eqnarray}\label{eq14}
H_{k_x}&=&\tau_1\begin{bmatrix}a_i^\dagger&d_i^\dagger\end{bmatrix}\hat{I}_\sigma\begin{bmatrix}b_i\\c_i\end{bmatrix}+\tau_2\begin{bmatrix}a_i^\dagger&d_i^\dagger\end{bmatrix}\hat{\sigma}_{z}\begin{bmatrix}a_{i+1}\\d_{i+1}\end{bmatrix}+\tau_2^*\begin{bmatrix}b_i^\dagger&c_i^\dagger\end{bmatrix}\hat{\sigma}_{z}\begin{bmatrix}b_{i+1}\\c_{i+1}\end{bmatrix}+\tau_4\begin{bmatrix}a_i^\dagger&d_i^\dagger\end{bmatrix}\hat{\sigma}_y\begin{bmatrix}b_i\\c_i\end{bmatrix}-\tau_4\begin{bmatrix}b_i^\dagger&c_i^\dagger\end{bmatrix}\hat{\sigma}_y\begin{bmatrix}a_{i+1}\\d_{i+1}\end{bmatrix}+H.C  \nonumber \\
&& +\tau_1\begin{bmatrix}a_i^\dagger&d_i^\dagger\end{bmatrix}(\hat{\sigma}_x\cos(k_x/2)-\hat{\sigma}_y\sin(k_x/2))\begin{bmatrix}a_i\\d_i\end{bmatrix}+\tau_1\begin{bmatrix}b_i^\dagger&c_i^\dagger\end{bmatrix}(\hat{\sigma}_x\cos(k_x/2)+\hat{\sigma}_y\sin(k_x/2))\begin{bmatrix}b_i\\c_i\end{bmatrix},
\end{eqnarray}
in which the new hopping parameters $\tau_i$ for this ladder geometry are defined as $\tau_1=t, \tau_2=i\lambda_{so}, \tau_3=te^{ik_x/2}, \tau_4=2\lambda_{so}\cos(k_x/2)$. 
Our effective Hamiltonian is now constructed.
However, for the reason which will become clear later, we make a unitary transformation with the following unitary matrix
$U=e^{i\frac{\pi}{4}\sigma_x}$ which satisfies $U\sigma_yU^T=\sigma_z$ where $T$ stands for transpose.
When applying this unitary transformation, we can obtain the resulting transformed operators $a^\prime, b^\prime, c^\prime$ and $d^\prime$ as follows:
\begin{eqnarray}\label{eq15}
\begin{bmatrix}a_i^\prime(c_i^\prime)\\b_i^\prime(d_i^\prime)\end{bmatrix}=U\begin{bmatrix}a_i(c_i)\\b_i(d_i)\end{bmatrix}=\frac{\sqrt{2}}{2}\begin{bmatrix}1&i\\i&1\end{bmatrix}\begin{bmatrix}a_i(c_i)\\b_i(d_i)\end{bmatrix}.
\end{eqnarray}

After performing the unitary transformation, it becomes straightforward to devide the Hamiltonian in Eq.~\eqref{eq14} 
into hopping term $H_0$ and on-site term $H_{on}$ as the following
\begin{eqnarray}\label{eq20}
\hat{H}=H_0+H_{on},
\end{eqnarray}
where
\begin{eqnarray}\label{HH0}
		H_0&=&\tau_1\begin{bmatrix}a_i^{\dagger\prime}&d_i^{\dagger\prime}\end{bmatrix}\hat{I}_\sigma\begin{bmatrix}b_i^\prime\\c_i^\prime\end{bmatrix}+\tau_2^\prime\begin{bmatrix}a_i^{\dagger\prime}&d_i^{\dagger\prime}\end{bmatrix}\hat{\sigma}_{y}\begin{bmatrix}a_{i+1}^\prime\\d_{i+1}^\prime\end{bmatrix}+(\tau_2^\prime)^*\begin{bmatrix}b_i^{\dagger\prime}&c_i^{\dagger\prime}\end{bmatrix}\hat{\sigma}_{y}\begin{bmatrix}b_{i+1}^\prime\\c_{i+1}^\prime\end{bmatrix}+\tau_4^\prime\begin{bmatrix}a_i^{\dagger\prime}&d_i^{\dagger\prime}\end{bmatrix}\hat{\sigma}_z\begin{bmatrix}b_i^\prime\\c_i^\prime\end{bmatrix} \nonumber \\
		&&-\tau_4^\prime\begin{bmatrix}b_i^{\dagger\prime}&c_i^{\dagger\prime}\end{bmatrix}\hat{\sigma}_z\begin{bmatrix}a_{i+1}^\prime\\d_{i+1}^\prime\end{bmatrix}+H.C+\tau_3^{\prime}\begin{bmatrix}a_i^{\dagger\prime}&d_i^{\dagger\prime}\end{bmatrix}\hat{\sigma}_x\begin{bmatrix}a_i^\prime\\d_i^\prime\end{bmatrix}+\tau_3^{\prime}\begin{bmatrix}b_i^{\dagger\prime}&c_i^{\dagger\prime}\end{bmatrix}\hat{\sigma}_x\begin{bmatrix}b_i^\prime\\c_i^\prime\end{bmatrix},
\end{eqnarray}
and
\begin{eqnarray}
H_{on}=-\Delta(k_x)\begin{bmatrix}a_i^{\dagger\prime}&d_i^{\dagger\prime}\end{bmatrix}\hat{\sigma}_z\begin{bmatrix}a_i^\prime\\d_i^\prime\end{bmatrix}+\Delta(k_x)\begin{bmatrix}b_i^{\dagger\prime}&c_i^{\dagger\prime}\end{bmatrix}\hat{\sigma}_z\begin{bmatrix}b_i^\prime\\c_i^\prime\end{bmatrix}.
\label{ME}
\end{eqnarray}	 
Here, the new hopping parameters of the transformed Hamiltonian are defined as $ \tau_2^\prime=\lambda_{so}, \tau_3^\prime=t\cos(k_x/2), \tau_4^\prime=\tau_4$ and the on-site energies $\Delta(k_x)=t\sin(k_x/2)$.
A graphical representation of the new parameters in the transformed Hamiltonian are shown in Fig~\ref{F1}(c).
In the upcoming sections, we will utilize the transformed Hamiltonian derived in this section to thoroughly investigate various properties of the system.

\section*{Results and discussion}\label{SECIII}

To present the findings of our study, this section is divided into three subsections, each focusing on a different aspect of the results. Firstly, we analyze the edge states of the system and discuss their properties for sufficiently wide ribbons. Secondly, we investigate the identification of the topological phase transition in the system by exclusively examining the properties of the edge state wave functions. Our investigation focuses on determining the conditions under which the system undergoes a transition, solely relying on the analysis of these edge states. Lastly, we delve into the intriguing phenomenon of bulk-edge correspondence violation in relatively small width ribbons and provide insights into its origin analitically. Through these subsections, we aim to provide a comprehensive understanding of the system's topological properties and shed light on its unique characteristics.

\subsection*{Edge stat analysis}
Our initial focus is on determining the edge states of the Hamiltonian~\eqref{eq20}. 
Considering that the presence of $H_{on}$ does not alter the edge state wave function~\cite{book}, we will confine ourselves to the edge states of $H_0$. Thus, our objective is to find $\vert\psi_{\text{edge}}\rangle$ that satisfies the following Schrödinger equation:
\begin{eqnarray}\label{eq22}
\hat{H_0}\vert\psi_{edge}\rangle=0
\end{eqnarray} 
The equation mentioned above holds when there exists a subspace on which the wave function vanishes. Our numerical analysis reveals that this subspace is formed by the basis states $a_i^\dagger\vert 0 \rangle$ and $c_i^\dagger\vert 0 \rangle$. This implies that our desired edge states have zero amplitudes on the lattice sites marked in red in Fig.~\ref{F1}(c). In other words, the wave function of the edge states vanishes at these particular lattice sites, that is
\begin{eqnarray}\label{eq23}
	\langle0\vert a_i\vert\psi_{edge}\rangle=0, \;\;\; \text{and} \;\;\; \langle0\vert c_i\vert\psi_{edge}\rangle=0.
\end{eqnarray}
Therefore, the general form of the edge state wave function is given by 
\begin{eqnarray}\label{WF}
\vert\psi_{edge}\rangle=\sum_{i=1}\alpha_{2i-1}\vert b_i\rangle+\alpha_{2i}\vert d_i\rangle,
\end{eqnarray}
where the coefficients $\alpha_i$ are the wave function amplitudes on $b$ and $d$ sites.  
By inserting the wave function Eq.~\eqref{WF} in  Eq.~\eqref{eq22} and choosing initial coefficients $\alpha_1$ and $\alpha_2$ one gets the following recursive relations:
\begin{eqnarray}\label{eqq}
\alpha_3&=&\frac{\tau_3^\prime}{\tau_2^\prime}\alpha_1+\frac{(\tau_1+\tau_4^\prime)}{\tau_2^\prime}\alpha_{2},\label{E15} \\
\alpha_4&=&\frac{(\tau_1-\tau_4^\prime)}{\tau_2^\prime}\alpha_{1}+\frac{\tau_3^\prime}{\tau_2^\prime}\alpha_2+\frac{(\tau_1+\tau_4^\prime)}{\tau_2^\prime}\alpha_{3},\label{E16} \\
\alpha_5&=&\alpha_{1}+\frac{(\tau_1-\tau_4^\prime)}{\tau_2^\prime}\alpha_{2}+\frac{\tau_3^\prime}{\tau_2^\prime}\alpha_3+\frac{(\tau_1+\tau_4^\prime)}{\tau_2^\prime}\alpha_{4}, \label{E17}  \\
\alpha_{i+4}&=&\beta_1\alpha_{i}+\beta_2\alpha_{i+1}+\beta_3\alpha_{i+2}+\beta_4\alpha_{i+3}, \;\;\; \text{for} \;\;\;  i\ge1, \label{E18} 
\end{eqnarray}
in which the coefficients $\beta_i$ reads as $\beta_1=1, \beta_2=\frac{(\tau_1-\tau_4^\prime)}{\tau_2^\prime}, \beta_3=\frac{\tau_3^\prime}{\tau_2^\prime},$ and $\beta_4=\frac{(\tau_1+\tau_4^\prime)}{\tau_2^\prime}$.

To obtain a solution that satisfies all of the aforementioned equations, we can start by introducing a general solution in the form:
\begin{eqnarray}\label{GS}
\alpha_{i+n}=x^n\alpha_i.
\end{eqnarray}
We then substitute this expression into Eq.~\eqref{E18}, which results in the following equation for $x$:
\begin{eqnarray}\label{eq35}
x^4-\beta_4x^3-\beta_3x^2-\beta_2x-\beta_1=0.
\end{eqnarray}
Our objective is to solve this equation to determine the proper values of $x$. Considering these solutions, we need to make a combination of them in a way that satisfies Eqs.~\eqref{E15} and Eq.~\eqref{E16} (Eq.~\eqref{E17} in not an independent equation.).
In principle, the fourth-order equation of Eq.~\eqref{E17} has four potential solutions. The largest solution of this equation, denoted as $x_4$, always satisfies the condition $x_4 > 1$ and should be excluded from the final combination due to its divergence~\cite{Num1}. Thus, our desired solution can be expressed in the following form:
\begin{eqnarray}\label{alphas}
\alpha_{1}&=&\frac{1}{\sqrt{\mathcal{N}}}(c_1+c_2+c_3)\nonumber\\
\alpha_{2}&=&\frac{1}{\sqrt{\mathcal{N}}}(c_1x_1+c_2x_2+c_3x_3)\nonumber\\
\alpha_{3}&=&\frac{1}{\sqrt{\mathcal{N}}}(c_1x_1^2+c_2x_2^2+c_3x_3^2)\nonumber\\ 
\alpha_{4}&=&\frac{1}{\sqrt{\mathcal{N}}}(c_1x_1^3+c_2x_2^3+c_3x_3^3 ),
\end{eqnarray}
in which $\mathcal{N}=\sum_{i,j}\frac{1}{1-x_ix_j}c_ic_j$ is the normalization factor and $c_i$ should be determined in such a way that satisfies the given conditions in Eqs.~\eqref{E15} and~\eqref{E16} which results in the following equations:
\begin{eqnarray}\label{S1}
\frac{c_1\beta_1}{x_1}+ \frac{c_2\beta_1}{x_2}+\frac{c_3\beta_1}{x_3}=0,
\end{eqnarray}
and
\begin{eqnarray}\label{S2}
c_1(\frac{\beta_2}{x_1}+\frac{\beta_1}{x_1^2})+c_2(\frac{\beta_2}{x_2}+\frac{\beta_1}{x_2^2})+c_3(\frac{\beta_2}{x_3}+\frac{\beta_1}{x_3^2})=0,
\end{eqnarray}
which should be solved with the additional condition
\begin{eqnarray}\label{S3}  
\frac{1}{\sqrt{\mathcal{N}}}(c_1+c_2+c_3)=1.
\end{eqnarray}

\begin{figure}
\centering
\includegraphics[width=0.6\linewidth]{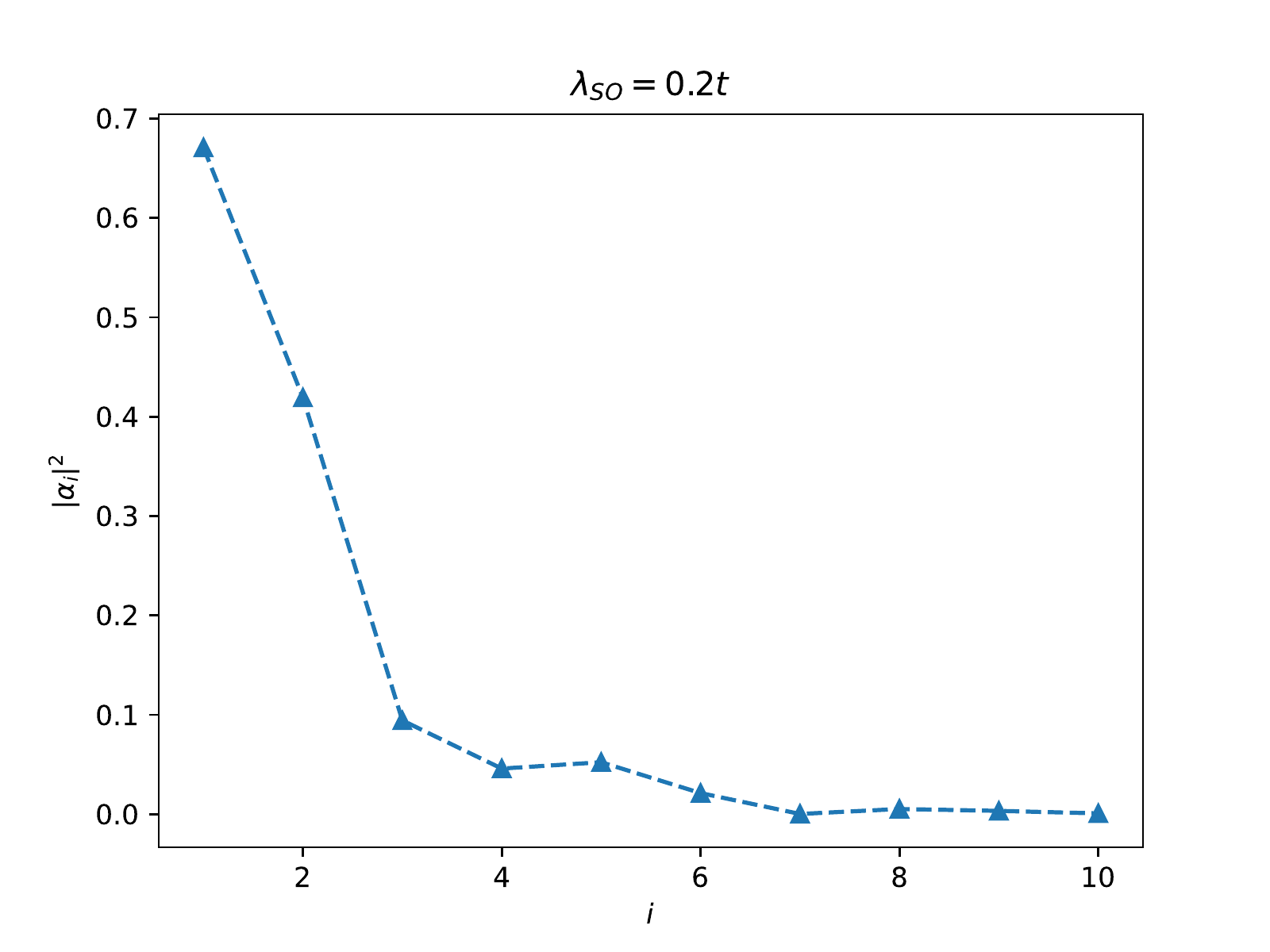} 
\caption{Plot of the amplitudes of the edge states wave function versus site index $i$ for the Kane-Mele model in the QSH phase. The wave function is calculated using Eq.~\eqref{WF} for $\lambda_{SO}=0.2t$ and $M=0$ at $k_x=0$.} 
\label{F3}
\end{figure}

Once the solution of Eqs.\eqref{S1}, \eqref{S2}, and \eqref{S3} is known\cite{Num2}, we can utilize it in Eq.\eqref{alphas} to obtain the edge state wave function as defined in Eq.\eqref{WF}. In Fig.~\ref{F3}, we have visually depicted this wave function for $\lambda=0.2 t_1$ and $k_x=0$.
The energy spectrum corresponding to this wave function can be obtained by evaluating~\cite{book} the expression $\langle \psi_{\text{edge}}\vert H_{\text{on}} \vert \psi_{\text{edge}} \rangle$, and it can be simplified as follows:
\begin{eqnarray}\label{Spectrum}
E(k_x)= t_1\sin(k_x/2),
\end{eqnarray}
This energy spectrum is depicted in Fig.~\ref{F2}, and it precisely coincides with the numerical edge band obtained for the same parameters. 

\subsection*{Identification of the topological transition}
In this section, our goal is to identify the topological transition point in our system by analyzing the edge state wave function that we obtained previously. To achieve this, we introduce the staggered on-site potential term in the Kane-Mele Hamiltonian of Eq.~\eqref{eq1}, which facilitates the crossing of the transition point.
In this regime ($M\neq 0$), the only modification is to adjust the hopping parameter $\tau_3^\prime$ as follows:
\begin{eqnarray}
\tau_3^\prime(M)\equiv  \tau_3^\prime+iM= t_1\cos(k_x/2)+iM. 
\end{eqnarray}
To obtain the new edge states in the presence of $M$, we need to repeat the analysis outlined in the previous subsection. The only difference is that now the recursion relation of Eq.~\eqref{E18} should be modified as follows:
\begin{eqnarray}
\alpha_{i+4}=\beta_1\alpha_{i}+\beta_2\alpha_{i+1}+\beta_3^\prime(M)\alpha_{i+2}+\beta_4\alpha_{i+3},
\end{eqnarray}
where the modified parameter $\beta_3^\prime(M)$ is given by:
\begin{eqnarray}
\beta_3^\prime(M)=\frac{\tau_3^\prime(M)}{\tau_2^\prime}.
\end{eqnarray}
By performing this analysis and considering the general solution of Eq.~\eqref{GS} we obtain the following equation for $x$:
\begin{eqnarray}\label{eq50}
x^4-\beta_4x^3-\beta_3^\prime (M) x^2-\beta_2x-\beta_1=0.
\end{eqnarray}
In our previous discussion, we noted that in the absence of $M$, one solution of the equation is always greater than one ($x_4 > 1$), while the remaining three solutions are not ($x_1, x_2, x_3 < 1$). However, the presence of $M$ introduces a change in the conditions, allowing for the possibility of another solution to be greater than one, depending on the value of $M$ (particularly for $M \gg t_1$ as shown in \cite{Book}). This additional solution is referred to as $x_3$. Consequently, under certain conditions, it is possible to have $x_1$ and $x_2$ less than one, while $x_3$ and $x_4$ are greater than one, rendering it impossible to find a solution for the edge states. This indicates that there exists a criterion which distinguishes between the regimes with and without a solution for the edge states. This criterion is therefore defined by $x_3=1$.
To determine which values of $M$ (which we call $M^*$) satisfy this criterion, we can consider $x_3=e^{i\theta}$ and substitute it into Eq.~\eqref{eq50} and after performing some simple algebra, this leads to the following equation:
\begin{eqnarray}
\lambda_{so}e^{2i\theta}-te^{i\theta}-2\lambda_{so}e^{i\theta}-te^{-i\theta}+2\lambda_{so}e^{-i\theta}-\lambda_{so}e^{-2i\theta}-(t+iM^*)=0.
\end{eqnarray}
If we equate the real and imaginary parts of the left side of the above equation to zero, we obtain the following equations 
\begin{eqnarray}
&2\cos(\theta)=-1\nonumber\\
&-4\lambda_{so}\sin(2\theta)+2\lambda_{so}\sin(2\theta)=M^*
\end{eqnarray}
which results in
\begin{eqnarray}
M^*=3\sqrt{3}\lambda_{so}.
\end{eqnarray}

Therefore, for values of $M$ less than $M^*$, a solution for the edge states can be obtained. However, when $M$ reaches the critical value $M^*$, the edge states become unstable and disappear, leading to the emergence of a gap in the spectrum at $k_x=0$.
This is a fascinating observation as it demonstrates that the transition point can be determined solely by analyzing the wave function of the edge states. By studying the behavior of the edge states, we were able to identify the critical value of $M$ where the transition occurs. This highlights the significance of edge state analysis in understanding the topological properties and phase transitions in the system.

Before ending this subsection, it is worth mentioning that the presence of $M$ does not alter the energy spectrum of the edge states while we are in the topological phase. This is due to the fact that the parameter $\Delta(k_x)$, which describes the on-site energies of the Hamiltonian $H_0$ in Eq.\eqref{ME}, remains unaffected by the presence of $M$. As a result, the energy spectrum of the edge states remains unchanged and can still be described by Eq.\eqref{Spectrum}.

\subsection*{Bulk-edge correspondence violation in relatively small width ribbons}
So far, our analysis has focused on sufficiently wide ribbons where the wave functions localized at the top and bottom edges, denoted as $\vert\psi_{\text{edge}}^{\text{top}}\rangle$ and $\vert\psi_{\text{edge}}^{\text{bottom}}\rangle$ respectively, do not overlap. However, it is an intriguing question to investigate the behavior when the ribbon width becomes small enough for these localized edge modes to overlap. For this purpose, we start by considering a simple SSH chain of length $N$ that can be described by the following tight-binding Hamiltonian,   
\begin{eqnarray}
\hat{H}_{SSH}=\upsilon a_{i}^{\dagger}b_{i}+\omega b_{i}^{\dagger}a_{i+1}+ H.C.
\end{eqnarray}
The corresponding end mode of this chain can be described as~\cite{book}: 
\begin{eqnarray}
\vert\psi_{L}\rangle=\sum_ia_i\vert a_i\rangle,
\end{eqnarray}
where $a_i=x_{SSH}^ia_1$ and $x_{SSH}=-\frac{\upsilon}{\omega}$ with $a_1=\frac{1}{\sqrt{\mathcal{N}_{SSH}}}$. Here, $\mathcal{N}_{SSH}$ is the normalization factor associated to the edge states of this SSH chain and the subscript $L$ refers to the left side of the chain.  
In a similar manner, the localized edge state at the right side of the chain is given by,
\begin{eqnarray}
\vert\psi_{R}\rangle=\sum_i b_i\vert b_i\rangle, \\
b_i=\sum_i x_{SSH}^{(N-i)}b_N.
\end{eqnarray}  
In the case where the length of the chain is finite, such that the amplitudes of the left edge mode near the right side of the chain are not close to zero, it can be demonstrated that there exists an energy gap denoted as $2\Delta^\prime$. The value of this mid gap energy can be expressed as follows:
\begin{eqnarray}\label{SSH-GAP}
\Delta^\prime=\langle\psi_{L}\vert \hat{H}_{SSH} \vert\psi_{R}\rangle=\frac{1}{\mathcal{N}_{SSH}}\upsilon\mid x_{SSH}\mid^N.
\end{eqnarray} 

\begin{figure}
\centering
\includegraphics[width=0.6\linewidth]{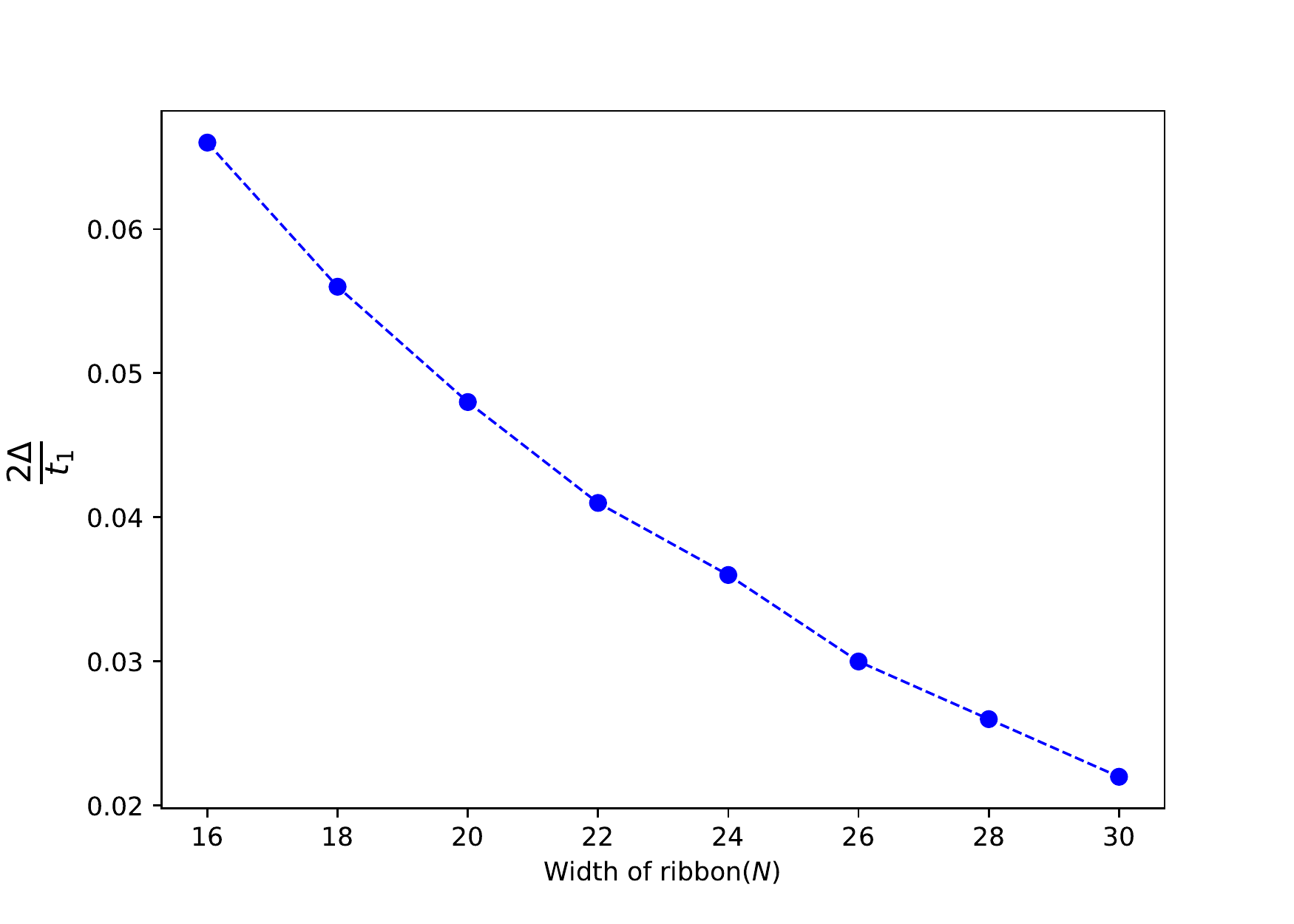} 
\caption{The behavior of the energy gap $\frac{2\Delta}{t_1}$ for a narrow ribbon versus the ribbon width $N$ at $k_x=0$. The analytical expression obtained in Eq.~\eqref{GG} reveals the dependence of the energy gap on the ribbon width, highlighting the impact of confinement effects on the edge bands of the system in the QSH phase.}
\label{F4}
\end{figure}

Let us now replace our SSH chain with the narrow ribbon of width $N$. So we need to consider  $\vert\psi_{\text{edge}}^{\text{top}}\rangle$ and $\vert\psi_{\text{edge}}^{\text{bottom}}\rangle$ instead of $\vert\psi_{L}\rangle$ and $\vert\psi_{R}\rangle$, that is:
\begin{eqnarray}
\vert\psi_{top}\rangle=\sum_i a_i\vert a_i\rangle \;\;\; \text{with}\;\;\;a_{i}=c_{1}\alpha_{1}^{i}+c_{2}\alpha_{2}^{i}+c_{3}\alpha_{3}^{i}
\end{eqnarray}
and
\begin{eqnarray}
\vert\psi_{bottom}\rangle=\sum_ib_i\vert b_i\rangle\;\;\; \text{with}\;\;\; b_{i}=c_{1}\alpha_{1}^{N-i}+c_{2}\alpha_{2}^{N-i}+c_{3}\alpha_{3}^{N-i}.
\end{eqnarray}
The only remain step is to replace $\hat{H}_{SSH}$ in Eq.~\eqref{SSH-GAP} with $H_0$ of Eq.~\eqref{HH0}  which results in the following band gap:
\begin{eqnarray}\label{GG}
2\Delta=2\langle\psi_{top}\vert H_0 \vert\psi_{bottom}\rangle \approx   \frac{1}{\mathcal{N}}4\tau_1c_{1}^2\mid x_{3}\mid^{2N}.
\end{eqnarray}
A graphical representation of this results is plotted in Fig.~\ref{F4} which shows the dependence of the energy gap  $2\Delta$ on the width of ribbon $N$ for $k_x=0$ and $\lambda_{SO}=0.2$.

Based on our analysis, it can be concluded that the bulk-edge correspondence, which states the presence of edge states in a topological system, does not hold for relatively narrow ribbons. In such cases, where the width of the ribbon is small enough for the edge modes to overlap, the system may not exhibit distinct edge states within the bulk energy gap. This deviation from the bulk-edge correspondence highlights the importance of considering the size and geometry of the system when studying topological properties which was studied before numerically~\cite{Kondo} and is addressed here analytically.


\begin{thebibliography}{}
\bibitem{Num1} For instance, for $t=1, \lambda_{so}=0.3$ and at $k_x=0$ they are $x_1=0.01208 + 0.51765i, x_2=0.01208 - 0.51765i, x_3= -0.62818,$  and $x_4=5.93735>1$.
\end{thebibliography}

\section*{Concluding remarks}\label{CONC}

In conclusion, our study focuses on the investigation of the edge states and their topological properties in the Kane-Mele model within a ribbon geometry with armchair boundaries. We have presented a novel analytical approach that utilizes a mapping procedure to transform the system into an extended SSH model, specifically a two-leg ladder, in the momentum space. This approach has allowed us to derive explicit expressions for the edge states, including their wave functions and energy dispersion. By analyzing the edge states, we have determined the condition for topological transition, providing insights into the unique characteristics of the edge states in the quantum spin Hall phase of the Kane-Mele model.

Furthermore, our study sheds light on the violation of the bulk-edge correspondence in relatively narrow ribbons. We have elucidated the underlying reasons for this violation, highlighting the importance of considering the size of the system in understanding its topological properties. The analytical properties obtained through our approach contribute to a comprehensive understanding of the edge states in the Kane-Mele model and provide valuable insights into the topological properties of such systems.

Overall, our research provides a new analytical framework for studying edge states in topological systems and contributes to the growing field of topological condensed matter physics. Further investigations can explore the implications of our findings in other topological models and the potential applications of edge states in quantum information processing and device design~\cite{Amini1, Amini3}.

\section*{Acknowledgements}

MA acknowledges the associateship award from the International Centre for Theoretical Physics (ICTP), Trieste, Italy. This support greatly facilitated the progress of the research and is greatly appreciated.
The first author would also like to express gratitude to the office of graduate studies at the University of Isfahan for their generous support and provision of research facilities. Their assistance has been instrumental in conducting this study.

\section*{Author contributions statement}

All of the authors have equally contributed to all aspects of the research presented here.

\end{document}